# Understanding the Magnetic Puzzles of Double Perovskites $A_2$FeOsO$_6$ ($A$=Ca, Sr)

Y. S. Hou, H. J. Xiang, and X. G. Gong

Key Laboratory of Computational Physical Sciences (Ministry of Education), State Key Laboratory of Surface Physics, and Department of Physics, Fudan University, Shanghai 200433, People's Republic of China

**Abstract**

Double perovskites Sr$_2$FeOsO$_6$ and Ca$_2$FeOsO$_6$ show puzzling magnetic properties, the former a low-temperature antiferromagnet while the later a high-temperature insulating ferrimagnet. Here, in order to understand the underlying mechanism, we have investigated the frustrated magnetism of $A_2$FeOsO$_6$ by employing density functional theory and maximally-localized Wannier functions. We find that lattice distortion enhances the antiferromagnetic nearest-neighboring Fe-O-Os interaction but weakens the antiferromagnetic interactions through the Os-O-O-Os and Fe-O-Os-O-Fe paths, which is responsible for the magnetic transition from the low-temperature antiferromagnetism to the high-temperature ferrimagnetism with the decrease of the radius of the $A^{2+}$ ions. We also discuss the $5d^3$-$3d^5$ superexchange and propose such superexchange is intrinsically antiferromagnetic instead of the expected ferromagnetic. Our work illustrate that the magnetic frustration can be effectively relieved by lattice distortion, which provides another dimension to tune the complex magnetism in other $3d$-$5d$ ($4d$) double perovskites.

**PACS numbers:** 75.50.Gg, 75.50.Ee, 71.20.-b, 71.27.+a

# I. INTRODUCTION

Double perovskite (DP) oxides $A_2BB'O_6$, where A is alkaline-earth or rare-earth metals and B($B'$) are transition metals (TMs), have attracted considerable attention due to the numerous interesting physical discoveries, including the room-temperature (RT) half-metallicity [1], high-temperature (HT) insulating ferrimagnetism [2, 3], multiferroicity [4], ferromagnetism [5] and so on. Especially, osmium-based DPs $Sr_2FeOsO_6$ (SFOO), $SrCaFeOsO_6$ (SCFOO) and $Ca_2FeOsO_6$ (CFOO) are intensively investigated recently because of their puzzling magnetic behaviors [3, 6-11].

Experimentally, it is showed that SFOO adopts two types of antiferromagnetism, namely, the AF1 and AF2, whose Neel temperatures are $T_N^{AF1} = 140\,K$ and $T_N^{AF2} = 67\,K$ [6], respectively. Theoretically, the mechanisms of the occurrence of the AF1 and AF2 are under debate. For the AF1, it was widely accepted that the ferrimagnetic (FIM) *ab* planes are coupled to the neighboring planes by a ferromagnetic (FM) Fe-O-Os superexchange [3, 8, 9]. However, it was recently suggested that these FIM *ab* planes may be coupled by the antiferromagnetic (AFM) Os-O-O-Os interactions [10]. For the AF2, Morrow *et al.* proposed that the long-range Fe-Fe AFM interaction through the four-bond Fe-O-Os-O-Fe path dominates and leads to the AFM-type Fe-Os chains along the *c* axis [10], while Kanungo *et al.* showed that the long-range Os-Os AFM interaction through the four-bond Os-O-Fe-O-Os path is responsible [9]. For the magnetic ordering temperature, there is no quantitative understanding why the $T_N$ of AF1 is low up to now.

Surprisingly, SCFOO is a ferrimagnet in comparison with the antiferromagnetic SFOO and has a higher magnetic ordering temperature $T_C \approx 210\,K$ [10] than SFOO, although the main lattice structure difference between SCFOO and SFOO is that the Fe-O-Os bond angle along the *c* axis in SCFOO is smaller than that in SFOO. More interestingly, CFOO is an insulating ferrimagnet with an even higher magnetic ordering temperature $T_C \approx 320\,K$ [3, 10], which could not be accounted for by the generalized double exchange mechanism [12]. To the best of our knowledge, there is still a lack of comprehensive and unified understandings on how the low-temperature (LT) antiferromagnetism of $A_2FeOsO_6$ transforms to the HT ferrimagnetism with the decrease of the radius of the $A^{2+}$ ions.

In this paper, to obtain a unified insight into all these puzzles, we have systematically investigated the frustrated magnetism of the DPs CFOO, SCFOO and SFOO by employing density functional theory (DFT) and maximally-localized Wannier functions (MLWFs). We find that lattice distortion enhances the AFM Fe-O-Os interaction but weakens the AFM interactions of the Os-O-O-Os and Fe-O-Os-O-Fe paths. As a result of the serious lattice distortion, therefore, CFOO has the strong and dominant AFM interaction between the nearest neighboring (NN) $Fe^{3+}$ and $Os^{5+}$ ions. Consequently, the NN $Fe^{3+}$ and $Os^{5+}$ ions are antiparallelly coupled and the experimentally observed ferrimagnetism is established. Simultaneously, its AFM interactions through the Os-O-O-Os and Fe-O-Os-O-Fe paths are weak, so the Os-Os and Fe-Fe magnetic frustration is effectively relieved and it has a very high $T_C$. Because SCFOO is less distorted compared to CFOO, its magnetic frustration becomes stronger. Accordingly, its $T_C$ is lowered. In the tetragonal $I4/m$ structure of SFOO, lattice distortion vanishes along the *c* axis but is very similar to that of CFOO in the *ab* plane. This special lattice distortion pattern results in the in-plane NN $Fe^{3+}$ and $Os^{5+}$ ions being antiparallelly aligned and the FM chains along the *c* axis. The resulting magnetic structure is the strongly frustrated antiferromagnetism AF1 with a low Neel temperature $T_N$. Lastly, strong spin-lattice coupling leads to the transformation from AF1 to AF2. Our work illustrates that the magnetic frustration can be effectively relieved by lattice distortion, which may be responsible for the complex magnetism in other *3d-5d* (*4d*) DPs.

## II. COMOPUTATIONAL DETAILS

First-principles calculations based on DFT are performed within the generalized gradient approximation (GGA) according to the Perdew-Burke-Ernzerhof (PBE) parameterization as implemented in Vienna *Ab initio* Simulation Package (VASP) [13]. The projector-augmented wave method [14], an energy cutoff of 500 eV and a gamma-centered k-point mesh grid are used. Ion positions are relaxed towards equilibrium with the Hellmann-Feynman forces on each ions less than 0.01 eV/Å. We use the simplified (rotationally invariant) coulomb-corrected density functional (DFT+U) method according to Dudarev *et al.* [15]. $U_{Fe}^{eff} = 4.0\,eV$, and $U_{Os}^{eff} = 2.0\,eV$ [9] are applied to the Fe *3d* and Os *5d* states, respectively. With these effective U, the calculated band gap of CFOO is about 1.19 eV, very close to the experimentally measured activation energy ($E_{gap} = 1.2\,\text{eV}$) [3]. It has been demonstrated [7] that spin-orbit coupling (SOC) in CFOO is insignificant, so SOC is not taken into account in the present work. Hopping integrals between 3$d$/5$d$ orbitals are extracted from the real-space Hamiltonian matrix elements in the non-spin-polarized MLWFs basis. MLWFs are obtained by employing the **vasp2wannier90** interface in combination with the **wannier90** tool [16]. In order to obtain the 3$d$/5$d$-like Wannier functions, we construct MLWFs in a suitable energy window mainly containing the 3$d$/5$d$ antibonding states. All MLWFs are considered to be well converged if the total spread change within 50 successive iterations is smaller than $10^{-9}\,\text{Å}^2$.

## III. RESULTS AND DISCUSSIONS

We first demonstrate the magnetic interaction between the NN $Fe^{3+}$ and $Os^{5+}$ ions through the Fe-O-Os path in CFOO is intrinsically AFM and that lattice distortion can effectively relieve the magnetic frustration and increase $T_c$. Then we show that the competing magnetic interactions in the tetragonal $I4/m$ structure of SFOO give rise to the frustrated AF1 antiferromagnetism with a low Neel temperature $T_N$. We give explanations to the ferrimagnetism of SCFOO with a lower $T_C$. Last, we propose an important and general rule on the $3d^5$-$5d^3$ superexchange, which will help one to understand the complex magnetism in other DPs.

### A. Intrinsically AFM interaction between the NN $Fe^{3+}$ and $Os^{5+}$ ions and the effect of lattice distortion on the ferrimagnetism in CFOO

The lattice of CFOO is severely distorted. It has the monoclinic structure with the space group of $P2_1/n$ [3]. Lattice distortion of DPs $A_2BB'O_6$, depending on the ionic radii of the constituent ions, can be characterized by the tolerance factor

$$t = \frac{r_A + r_O}{\sqrt{2} \times \left[0.5 \times \left(r_B + r_{B'}\right) + r_O\right]}$$

where $r_A$, $r_B$, $r_{B'}$ and $r_O$ are the ionic radii of the respective ions. The tolerance factor of CFOO ($t_{CFOO} = 0.9639$) is smaller than 1. Accordingly, both FeO$_6$ and OsO$_6$ octahedra are severely tilted and rotated (see inset of FIG.1). The experimental out-of-plane and in-plane Fe-O-Os angles are 151.5° and 154.0°, respectively. After the structure relaxation, these two kinds of angles become 149.7° and 149.1°.

Lattice distortion has been suggested to cause the ferrimagnetism in CFOO [3]. According to Goodenough-Kanamori rules [17], the superexchange between the NN $Fe^{3+}(t_{2g}^3 e_g^2)$ and $Os^{5+}(t_{2g}^3)$ ions is expected to be FM. However, experimental measurements show that CFOO is FIM [3], indicating that the superexchange between the NN $Fe^{3+}$ and $Os^{5+}$ ions is AFM. Feng et al. [3] proposed this ferrimagnetism was driven by the lattice distortion, since $d^3$-$d^5$ magnetic interaction switched from FM to AFM in a range of 125°-150° and the Fe-O-Os angles in CFOO were just in the vicinity of this range. Their proposal is based on such underlying assumptions: (1) the AFM interaction due to $t_{2g}$ electrons is (much) weaker than the FM interaction due to the $e_g$ electrons; (2) the former has a weaker dependence on the Fe-O-Os angle than the latter [18]. However, it is highly possible that the $5d$ $t_{2g}$ electrons of the $Os^{3+}$ ions are spatially extended so that the former is stronger than the latter, which leads to an AFM interaction between the NN $Fe^{3+}$ and $Os^{5+}$ ions.

In order to unveil why CFOO is FIM and how lattice distortion affects this ferrimagnetism, we have systematically explored the effect of lattice distortion on its magnetic interaction. Since the positions of $O^{2-}$ ions are vital to the magnetic interaction, we perform series of calculations using a linear superposition of the Wyckoff positions of the relaxed structure and the pseudo-cubic one in which the oxygen atoms are artificially positioned to make Fe-O-Os angles straight. The lattice constants and the positions of the $Fe^{3+}$, $Os^{5+}$ and $Ca^{2+}$ ions are fixed. The atomic positions can be computed as

$$\boldsymbol{R}(\alpha_x) = (1-\alpha_x)\boldsymbol{R}^{relax} + \alpha_x \boldsymbol{R}^{cubic}$$

where $\boldsymbol{R}^{relax}$ and $\boldsymbol{R}^{cubic}$ are the position vectors of the relaxed and pseudo-cubic structures, respectively, and $\alpha_x$ varies between 0 and 1. For example, $\alpha_x = 0$ corresponds to the relaxed structure and $\alpha_x = 1$ to the pseudo-cubic one. Thus $\alpha_x$ characterizes the lattice distortion induced by $O^{2-}$ ions. The dominant magnetic interactions are divided into three groups shown in the inset of FIG. 1. The first group is the superexchange between the NN $Fe^{3+}$ and $Os^{5+}$ ions. The second one is the super-superexchange between the next near-neighboring (NNN) $Os^{5+}$ ions. The third one is the long-range Fe-Fe interaction through the four-bond Fe-O-Os-O-Fe path. Technically, we adopt the four-states mapping method [19] to evaluate these magnetic interactions. Note that a positive exchange constant $J$ corresponds to the AFM interaction whereas a negative one to the FM interaction.

We find the magnetic interaction between the NN $Fe^{3+}$ and $Os^{5+}$ ions is intrinsically AFM. The calculated magnetic exchange constants of Fe-O-Os paths in the pseudo-cubic structure are shown in the FIG. 1a. They are all positive and thus AFM. The intrinsically AFM interaction of the Fe-O-Os path can be qualitatively understood based on the extended Kugel-Khomskii model [20-22]. According to this model, magnetic interactions can be evaluated based on the hopping integrals and on-site energies, namely,

$$J_{ij} = J_{ij}^{AFM} + J_{ij}^{FM} = \sum_{afm} \frac{4 t_{ij}^{afm} \cdot t_{ij}^{afm}}{(U + \Delta_{ij})} - \sum_{fm} \frac{4 t_{ij}^{fm} \cdot t_{ij}^{fm} \cdot J_H}{(U + \Delta_{ij})(U + \Delta_{ij} - J_H)}$$

where U, $J_H$ and $\Delta_{ij}$ are the on-site Coulomb interaction, the Hund's coupling and the energy difference between the $i^{th}$ and $j^{th}$ energy levels, respectively. $t_{ij}^{afm}$ ($t_{ij}^{fm}$) is the hopping integral. The first term in $J_{ij}$ describes the AFM contribution due to the hybridization between two occupied orbitals. The second term describes the FM contribution due to the hybridization between the occupied and empty orbitals. In order to elucidate why the magnetic interaction between the NN $Fe^{3+}$ and $Os^{5+}$ ions is intrinsically AFM, we take the Fe-O-Os path along the $c$ axis in the pseudo-cubic CFOO as a typical example. Its detailed hopping integrals and energy levels are given in the right panel of FIG. S1 of supplemental material (SM) [23]. Compared with the FM interaction between the NN $Mn^{3+}$ ions in the cubic $LaMnO_3$ (LMO) [24], two pivotal factors drive the magnetic interaction between the NN $Fe^{3+}$ and $Os^{5+}$ ions in the pseudo-cubic CFOO to be intrinsically AFM. The first one is the very large energy difference $\Delta$ (up to 3.0 eV) between the occupied $e_g$ orbitals of $Fe^{3+}$ ion and the unoccupied one of $Os^{5+}$ ion. This gives weak FM contribution. The second one is the rather large hopping integrals between the occupied $t_{2g}$ orbitals of $Fe^{3+}$ and $Os^{5+}$ ions. For instance, the leading hopping integral is 0.27 eV. This gives strong AFM contribution. Therefore AFM contribution dominates over the FM one, giving rise to the intrinsically AFM interaction between the NN $Fe^{3+}$ and $Os^{5+}$ ions.

In addition, we find that lattice distortion can effectively relieve the magnetic frustration in CFOO, thereby raising the FIM phase transition temperature $T_C$. Shown in the FIG. 1a and 1b are the evolutions of the dominant magnetic exchange constants as $\alpha_x$ increases. As lattice distortion is gradually becoming serious, the NN AFM interactions between the NN $Fe^{3+}$ and $Os^{5+}$ ions slightly enhance, but the NNN AFM interactions between the NNN $Os^{5+}$ ions dramatically weaken. Simultaneously, the long-range four-bond Fe-O-Os-O-Fe AFM interactions along the pseudo-cubic [001], [010] and [100] axes also become weaker (see Fig. 1c). Since $Os^{5+}$ ions form a face-centered sublattice with geometrically frustrated edge-sharing tetrahedrons, antiferromagnetically interacting $Os^{5+}$ ions are strongly frustrated. On one hand, the dominant NN Fe-O-Os AFM interactions require the magnetic moments of $Fe^{3+}$ ions to be parallelly aligned. On the other hand, the AFM magnetic interaction through the four-bond Fe-O-Os-O-Fe paths requires the magnetic moments of $Fe^{3+}$ ions to be antiparallelly aligned. Obviously, $Fe^{3+}$ ions are magnetically frustrated. Because lattice distortion weakens the NNN Os-O-O-Os and long-range four-bond Fe-O-Os-O-Fe AFM interactions, it can effectively relieve the Os-Os and Fe-Fe magnetic frustration. Accompanied with such frustration relieving is the raising of the FIM phase transition temperature $T_C$. FIG. S2 of SM shows the evolution of $T_C$ obtained by Monte Carlo (MC) as lattice distortion weakens. It clearly shows that the $T_C$ of the relaxed structure ($\alpha_x=0$) (about 266K, close to the experimentally measured one ($T_C \approx 320$K) [3]) is higher than that of the less distorted one ($\alpha_x=0.25$). Note that $T_C$ slightly increases along with the weakening of lattice distortion for large $\alpha_x$. This is because the magnetic ground state of CFOO with small lattice distortion (see the inset of FIG. 1c) is no longer FIM but AFM with the AF1 order as appearing in the SFOO.

FIG. 2a demonstrates the mechanism that lattice distortion enhances the NN Fe-O-Os AFM interaction. For the illustration purpose, we take the Fe-O-Os path along the $c$ axis as an example. FIG. S1 of SM shows the detailed leading hopping integrals and energy levels in the relaxed and pseudo-cubic structures, respectively. These hopping integrals clearly indicates that lattice distortion tremendously reduces the electron hopping between the occupied $e_g$ orbitals of $Fe^{3+}$ ions and the unoccupied one of $Os^{5+}$ ions. Consequently, one can conclude based on the formula of

$J_{ij}$ that lattice distortion extraordinarily reduces the FM contribution to the NN superexchange. In contrast, lattice distortion has a rather minor effect on the AFM contribution, because it brings into considerable hoppings between the occupied $e_g$ orbitals of $Fe^{3+}$ ions and the occupied $t_{2g}$ orbitals of $Os^{5+}$ ions, although it reduces the hopping between the occupied $t_{2g}$ orbitals of $Fe^{3+}$ and $Os^{5+}$ ions as well. Therefore, lattice distortion enhances the NN AFM interaction by both dramatically reducing the FM contribution and maintaining the AFM contribution almost unchanged.

We find that the NNN AFM interaction between the NNN $Os^{5+}$ ions is weakened by lattice distortion. This is because such NNN super-superexchange has sensitive dependence on the geometry of the Os-O-O-Os path. Shown in the insets of Fig. 2b are the geometries of the concerned Os-O-O-Os paths in the relaxed and pseudo-cubic structures. The detailed leading hopping integrals and energy levels between the investigated $Os^{5+}$ ions are shown in the FIG. S3 of SM. Note that FM contribution to the NNN super-superexchange is rather weak in the relaxed and pseudo-cubic structures because of small hopping integrals and large energy differences. So the NNN super-superexchange is basically determined by the AFM contribution. Comparing the two investigated Os-O-O-Os paths (FIG. 2b) clearly indicates that lattice distortion increases the O-O bond length. Such increase reduces the hopping between the $t_{2g}$ electrons of $Os^{3+}$ ions, as verified by the reduced hopping integrals from the pseudo-cubic structure to the relaxed one (see FIG. S3 of SM). Therefore lattice distortion blocks the $t_{2g}$ electron hopping through Os-O-O-Os path, thereby weakening the NNN AFM interaction.

### B. Competing magnetic interactions lead to the AF1 antiferromagnetism with a low $T_N$ in SFOO

SFOO adopts two different magnetic and lattice structures depending on temperature [6]. With temperature decreasing, its magnetic structure transforms from AF1 into AF2 antiferromagnetism and its lattice structure transforms from $I4/m$ into $I4$ with a dimerization between the NN $Fe^{3+}$ and $Os^{5+}$ ions along $c$ axis. In both AF1 and AF2, moments of $Fe^{3+}$ and $Os^{5+}$ ions are antiparallelly coupled in the $ab$ plane

(FIG. 3a and FIG. 3c). Spins order as ++++ along the $c$ axis in AF1 (FIG. 3b) but ++−−++−− in AF2 (FIG. 3d). The experimentally measured out-of-plane and in-plane Fe-O-Os angles in the $I4/m$-AF1 phase are $180.0°$ and $166.1°$ [8], respectively. They become $180.0°$ and $161.2°$ after structure relaxations.

Our study on the $I4/m$-AF1 phase (FIG. 4a and FIG. 4b) shows the out-of-plane NN AFM interaction ($J^2_{\text{Fe-Os}}$) is much weaker than the in-plane one ($J^1_{\text{Fe-Os}}$) and that the out-of-plane NNN AFM interaction ($J^2_{\text{Os-Os}}$) is stronger than the in-plane one ($J^1_{\text{Os-Os}}$), which are understandable based on our above results on CFOO. Because the Fe-O-Os angle along the $c$ axis is $180.0°$, similar to that in the pseudo-cubic CFOO, the out-of-plane NN AFM interaction $J^2_{\text{Fe-Os}}$ is weak. In the tetragonal $ab$ plane, lattice distortion is similar to that in the relaxed CFOO, so the in-plane NN AFM interaction $J^1_{\text{Fe-Os}}$ is strong. The weak in-plane NNN AFM interaction $J^1_{\text{Os-Os}}$ is due to the strong in-plane lattice distortion blocking the $Os^{5+}$ ions' $t_{2g}$ electron hopping, similar to the weak NNN AFM interactions in the relaxed CFOO. Besides, the out-of-plane NNN AFM interaction $J^2_{\text{Os-Os}}$ is stronger than that of the relaxed CFOO but weaker than that of the pseudo-cubic CFOO. Overall, such magnitude of the spin interactions is a result of the combination of the absence of the lattice distortion along the $c$ axis and strong lattice distortion in the tetragonal $ab$ plane. Finally, it is expected that the long-range four-bond Fe-O-Os-O-Fe AFM interaction along the $c$ axis ($J^3_{\text{Fe-Fe}} = 6.2\,\text{meV}$) is stronger than the in-plane one ($J^2_{\text{Fe-Fe}} = 5.1\,\text{meV}$).

Here we demonstrate how the competing magnetic interactions establish the AF1 in the tetragonal $I4/m$ structure of SFOO. First of all, it should be noted that the easy axis is the $c$ axis [6], that is, magnetic moments can only point up and down along the $c$ axis. Because the in-plane NN AFM interaction $J^1_{\text{Fe-Os}}$ is approximately four times of the magnitude of the in-plane NNN AFM interaction $J^1_{\text{Os-Os}}$ and the in-plane long-range four-bond Fe-O-Os-O-Fe AFM interaction $J^2_{\text{Fe-Fe}}$ and their pairwise numbers (Z's) are all the same ($Z = 4$), the optimal configuration is that, in the $ab$ plane, magnetic moments of $Fe^{3+}$ and $Os^{5+}$ ions are antiparallelly aligned, as observed in the experiments [6]. Besides, the out-of-plane NN, NNN and the long-range four-bond magnetic interactions are all AFM. If only the out-of-plane NN AFM interaction is taken into consideration, the FIM $Fe^{3+}$-$Os^{5+}$ layers should be antiparallelly coupled along the $c$ axis. In this case, the resulting magnetic structure is the FIM (FIG. 4a). If the out-of-plane NNN AFM interaction is taken into consideration, the FIM $Fe^{3+}$-$Os^{5+}$ layers should be parallelly coupled along the $c$ axis. In this case, the resulting magnetic structure is the AF1 (FIG. 4b), namely, the experimentally observed one. Finally, if only the long-range four-bond Fe-O-Os-O-Fe AFM interaction is taken into consideration, it gives rise to the AF2 (FIG. 4c). Obviously, the out-of-plane NN, NNN and long-range four-bond AFM interactions compete against each other to give rise to the different magnetic ground states. Since their magnitudes are comparable (see FIG. 3b), their pairwise numbers are the decisive factor in determining the optimal magnetic structure. Their pairwise numbers are $Z_{NN}=2$, $Z_{NNN}=8$ and $Z_{Fe-O-Os-O-Fe}=2$, respectively. This indicates that the out-of-plane NNN AFM interactions overwhelmingly dominate over the out-of-plane NN and long-range four-bond AFM interactions. Therefore the optimal magnetic configuration is the AF1.

The foregoing deduction can be further confirmed as follows. In the FIM, all the out-of-plane Fe-Fe and Os-Os pairs are frustrated (see FIG. 4a). In the AF1, all the out-of-plane Fe-Os and Fe-Fe pair are frustrated (see FIG. 4b). In the AF2, half of the out-of-plane Fe-Os and Os-Os pair are frustrated (see FIG. 4c). In terms of the out-of-plane NN, NNN and long-range four-bond AFM interactions, therefore, the formula-unit (f.u.) magnetic energies of the FIM, AF1 and AF2 are

$$E_{FIM}(J) = -2J^2_{Fe-Os} + J^3_{Fe-Fe} + 4J^2_{Os-Os} = 19.8\,\text{meV/f.u.}$$

$$E_{AF1}(J) = 2J^2_{Fe-Os} + J^3_{Fe-Fe} - 4J^2_{Os-Os} = -7.4\,\text{meV/f.u.}$$

$$E_{AF2}(J) = -J^3_{Fe-Fe} = -6.2\,\text{meV/f.u.}.$$

This indicates that FIM should have the highest energy, AF2 the media one and AF1 the lowers one. Such estimation is in accord with our DFT calculations: $E_{FIM}=-34.139\,\text{meV/f.u.} > E_{AF2}=-31.153\,\text{meV/f.u.} > E_{AF1}=-34.156\,\text{meV/f.u.}$. So the AF1 is established to relieve the magnetic frustration as much as possible.

The low Neel temperature $T_N$ of the AF1 is a result of the strong magnetic frustration. Actually, only the in-plane $J^1_{Fe-Os}$ and the out-of-plane $J^2_{Os-Os}$ AFM interactions are not frustrated in the AF1. However, the in-plane $J^1_{Os-Os}$, out-of-plane $J^2_{Fe-Os}$ and the long-range four-bond $J^2_{Fe-Fe}$, $J^3_{Fe-Fe}$ AFM interactions violate the AF1 antiferromagnetism and thus induce frustration. Our MC simulations indicate that, if only $J_{\text{Fe-Os}}^{1}$ and $J_{\text{Os-Os}}^{2}$ are taken into considerations, the $T_N$ of the AF1 is very high, up to 354K, sharply contradicting with the experimentally observe one (140K). To determine why the experimentally measured $T_N$ is so low, we performed four additional MC simulations: one with the in-plane $J_{\text{Os-Os}}^{1}$, one with the out-of-plane $J_{\text{Fe-Os}}^{2}$, one with the long-range four-bond $J_{\text{Fe-Fe}}^{2}$ and $J_{\text{Fe-Fe}}^{3}$, and one with all these magnetic interactions. The resulting specific heat versus temperature plots are presented in FIG 4d. As can be seen, $J_{\text{Os-Os}}^{1}$, $J_{\text{Fe-Os}}^{2}$ and the long-range four-bond $J_{\text{Fe-Fe}}^{2}$ and $J_{\text{Fe-Fe}}^{3}$ can all lower the $T_N$ for they are all frustrated. Moreover, the out-of-plane Fe-O-Os AFM interactions make the most contribution to the lowering of $T_N$ of the AF1. If all the dominating magnetic interactions are taken into consideration, the MC simulated $T_N$ is 155K, very close to the experimental one.

By comparing the magnetic exchange constants of $I4$ structure with that of $I4/m$ structure (see FIG. 3), one can find that the magnetic interactions situation in the former is very similar to that in the latter, with the exception that the rather slight dimerization along the $c$ axis in the former prominently enhances the out-of-plane NN AFM interactions $J_{\text{Fe-Os}}^{3}$ (see FIG. 3d), which indicates a very strong spin-lattice coupling. Like in the $I4/m$ structure, the long-range four-bond Fe-O-Os-O-Fe AFM interaction $J_{\text{Fe-Fe}}^{3}$ favors the formation of the AF2. Furthermore, the enhancement of the out-of-plane $J_{\text{Fe-Os}}^{3}$ favors this formation. Thus we attribute the AF2 antiferromagnetism in $I4$ structure to the strong spin-lattice coupling.

### C. Explanations to the ferrimagnetism of SCFOO with a lower $T_C$

Comparing SCFOO with SFOO and CFOO, one can conclude that its mediate lattice distortion causes its ferrimagnetism with a lower $T_C$. Experiments show that SCFOO has a rather similar lattice structure to that of CFOO [10]. However, its Fe-O-Os bond angles approach a more linear geometry than that of CFOO, because half of $Ca^{2+}$ ions are replaced by larger $Sr^{2+}$ ions. So it can be inferred that SCFOO should be still ferrimagnetic. Our detailed study shows that SCFOO is indeed FIM (see Part IV of SM), consistent with experiments [10]. Since its Fe-O-Os bond become more linear, its NN AFM interactions become weaker but its NNN

Os-O-O-Os and long-range four-bond Fe-O-Os-O-Fe AFM interactions become stronger. This is verified by the calculated magnetic exchange parameters, as listed in the Table *II* of SM. Consequently, its magnetic frustration gets stronger and its $T_C$ should be lowered. Our MC simulated $T_C$ of SCFOO is about 100K, lower than the corresponding $T_C$ (266K) of CFOO, consistent with the experimental observations [6, 10].

### D. Discussions on the $3d^5$-$5d^3$ superexchange

Based on the present work, an important and general rule on the $3d^5$-$5d^3$ superexchange in the DPs can be proposed as follows. It is generally accepted that the $d^5$-$d^3$ superexchange changes from FM for $\alpha > \alpha_c$ to AFM for $\alpha < \alpha_c$ with $135^\circ < \alpha_c < 150^\circ$ [17]. However, we demonstrate that the magnetic interaction between $3d^5$ and $5d^3$ TMs would be intrinsically AFM (This conclusion is independent of the choice of a reasonable U, see Table *III* of SM) and that this AFM interaction would enhance as its angle $\alpha$ decreases, as evidenced by the Fe-O-Os interactions in CFOO, SCFOO and SFOO. This intrinsically AFM interaction results from both the large hopping integrals between the occupied $t_{2g}$ orbitals and the large energy difference between the occupied $e_g$ orbitals of $3d$ TM and the unoccupied ones of $5d$ TM, for the former gives rise to strong AFM contribution and the latter gives rise to weak FM contribution to the $3d^5$-$5d^3$ superexchange. As the angle $\alpha$ decreases, the electron hoppings between the occupied $e_g$ orbitals of the $3d^5$ TM and the unoccupied ones of the $5d^3$ TM will substantially reduce but the electron hoppings between the occupied orbitals of $3d^5$ TM and $5d^3$ TM are maintained almost unchanged. Thus decreasing the angle $\alpha$ means reducing the FM contribution but maintaining the AFM contribution almost unchanged. Consequently, the AFM interaction of the $3d^5$-O-$5d^3$ path enhances with $\alpha$ decreasing.

### IV. SUMMARY

In conclusion, we have investigated the effect of lattice distortion on the frustrated magnetism of double perovskites $Ca_2FeOsO_6$, $Sr_2FeOsO_6$ and $Sr_2CrOsO_6$. We find that lattice distortion enhances the NN AFM Fe-O-Os interactions but weakens the AFM interactions of the Os-O-O-Os and Fe-O-Os-O-Fe paths. As lattice distortion gets more and more severe from $Sr_2FeOsO_6$ to $SrCaFeOsO_6$ and then to $Ca_2FeOsO_6$, therefore, the NN AFM Fe-O-Os interactions become stronger but the AFM interactions of Os-O-O-Os and Fe-O-Os-O-Fe paths become weaker. Consequently, the magnetic ground state transforms from antiferromagnetism to ferrimagnetism, and the magnetic transition temperature increases. We propose the $5d^3$-$3d^5$ superexchange is intrinsically antiferromagnetic instead of the expected ferromagnetic. Our work illustrates that the magnetic frustration can be effectively relieved by the lattice distortion in $3d$-$5d$ ($4d$) double perovskites.

## ACKNOWLEDGMENTS

Work was partially supported by NSFC, the Special Funds for Major State Basic Research, Foundation for the Author of National Excellent Doctoral Dissertation of China, The Program for Professor of Special Appointment at Shanghai Institutions of Higher Learning, Research Program of Shanghai municipality and the Ministry of Education, and Fok Ying Tung Education Foundation. We thank Dr. Jihui Yang, Prof. Xiao Gu, Prof. Zhixin Guo and Dr. Haiyuan Cao for useful discussions.

**Figures**

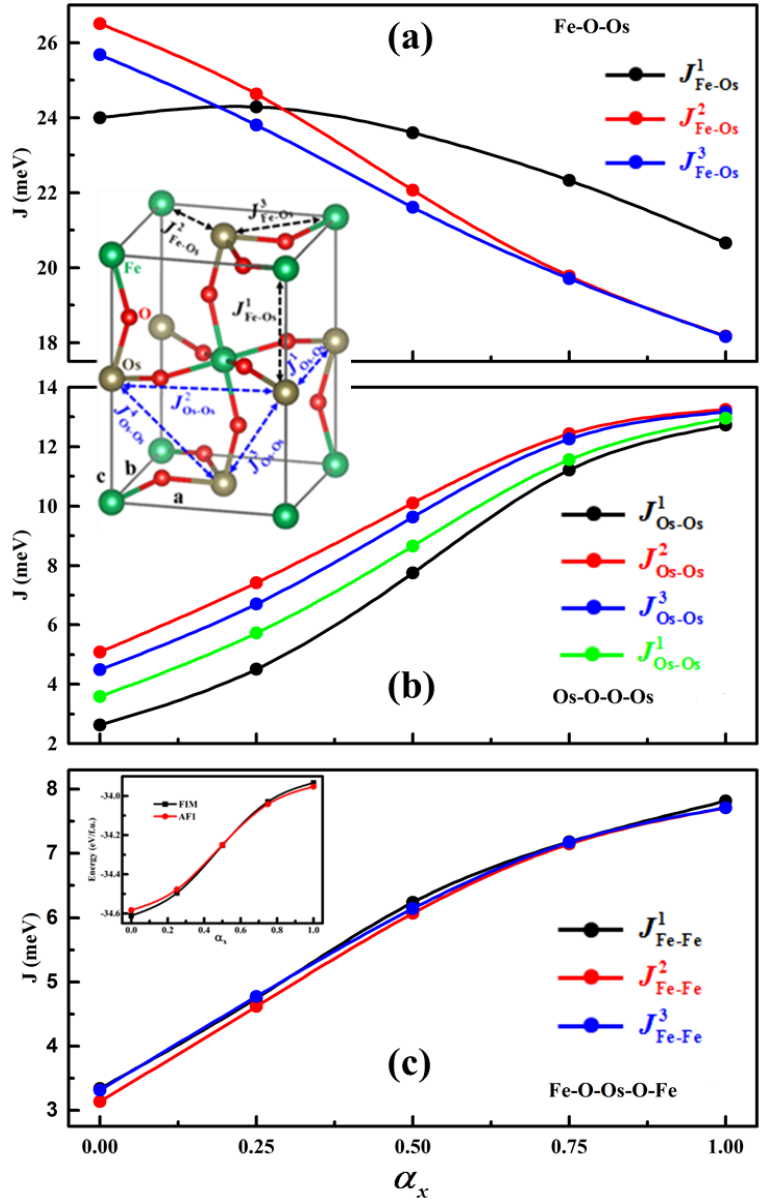

FIG. 1. (Color online) The dependence of the calculated magnetic parameters on $\alpha_x$ (lattice distortion). (a), (b) and (c) correspond to the NN Fe-O-Os superexchanges, NNN Os-O-O-Os super-superexchanges and the long-range four-bond Fe-O-Os-O-Fe interactions, respectively. The inset between (a) and (b) illustrates the crystal structure of $Ca_2FeOsO_6$ and the dominant magnetic interaction paths. The NN superexchange paths Fe-O-Os and NNN super-superexchange paths Os-O-O-Os are shown by the black and blue solid lines with double arrowheads, respectively. The long-range four-bond exchange paths $J^1_{Fe-Fe}$, $J^2_{Fe-Fe}$ and $J^3_{Fe-Fe}$ (not shown) are along the pseudo-cubic [001], [100] and [010] axes. $Ca^{2+}$ ions are omitted from the crystal structure for clarity. The inset in (c) is the dependence on $\alpha_x$ of the energy of the FIM (black) and AF1 (red) magnetic structures.

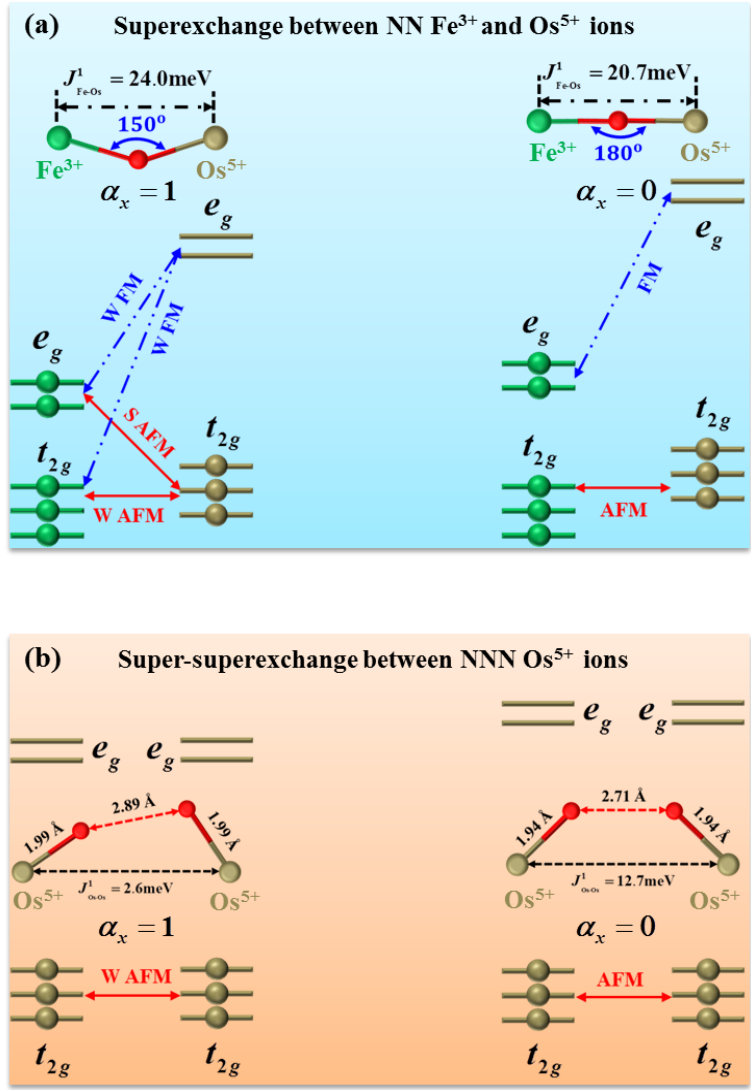

FIG. 2. (Color online) The mechanism that lattice distortion enhances the NN AFM interaction between $Fe^{3+}$ and $Os^{5+}$ ions (a) and weakens the NNN AFM interaction between $Os^{5+}$ ions (b) in $Ca_2FeOsO_6$. Solid (dashed) lines with double arrowheads indicate the electron hopping making AFM (FM) contribution to the NN superexchange or NNN super-superexchange. S and W represent "strong" and "weak", respectively. In (a), the FM contribution of $a_x = 1$ is weaker than that of $a_x = 0$. However, the AFM contribution of $a_x = 1$ is stronger than that of $a_x = 0$. In (b), the AFM contribution of $a_x = 1$ is weaker than that of $a_x = 0$. Insets in (a) and (b) are the local structures of Fe-O-Os and Os-O-O-Os paths, respectively. The relevant bond angles, bond lengths and calculated magnetic exchange constants are given.

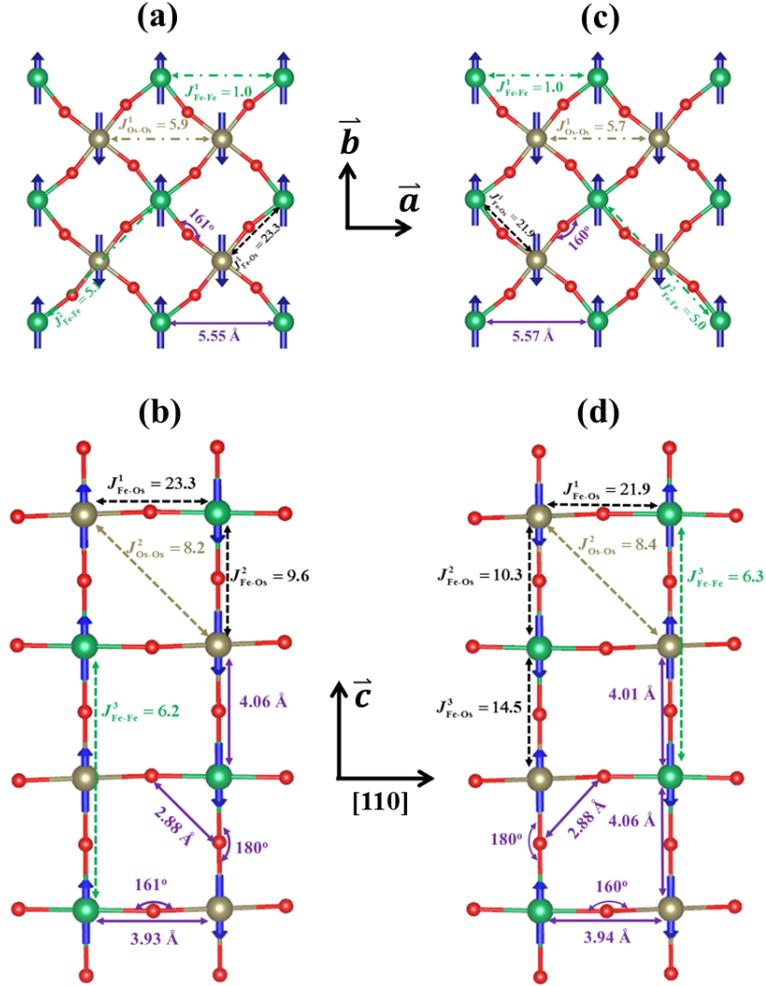

FIG.3. (Color online) The dominant magnetic exchange paths and the AF1, AF2 magnetic structures of $Sr_2FeOsO_6$. All magnetic exchange constants $J$ are in unit of meV. (a) and (b) correspond to the $I4/m$-AF1 phase. (c) and (d) correspond to the $I4$-AF2 phase. (a) and (c) are the spin arrangement in the tetragonal $ab$ plane. (b) and (d) are the spin ordering along the $c$ axis. Black arrows represent spins. Given are the relevant bond distances and angels obtained from DFT calculations.

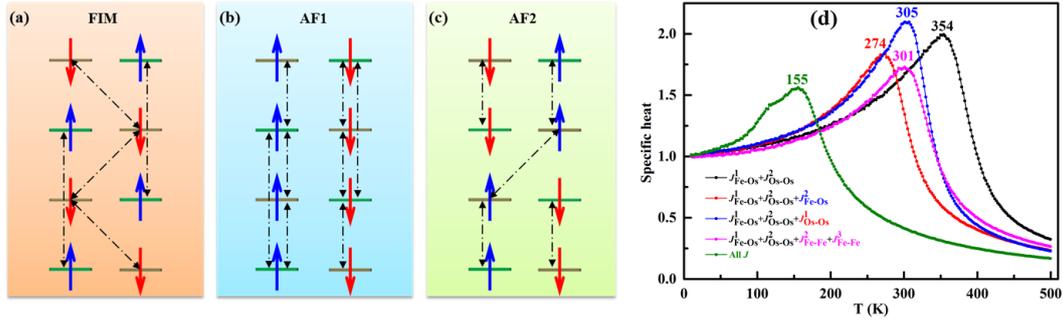

FIG. 4. (Color online) Shown are the FIM (a), AF1 (b) and AF2(c) magnetic structures. In (a), (b) and (c), the frustrated magnetic ions pairs are connected by the black dashed lines with double arrowheads. Fe (Os) sites are represented by the green (gray) horizontal lines. Blue (red) arrows represent up (down) spins. (d) is the specific heat of SFOO, calculated as a function of temperature T in terms of the spin exchange interactions.